\documentclass[preprintnumbers, floatfix,letterpaper,aps,prd,epsfig,nofootinbib, longbibliography, twocolumn]{revtex4-1}
\usepackage{bm,graphicx,dcolumn,epstopdf,epsf, latexsym,mathbbol, amssymb,amsmath, color, slashed, mathrsfs,mathcomp,simplewick}
\usepackage[center]{subfigure}
\usepackage{multirow}
\usepackage{makecell}
\usepackage{ytableau}
\usepackage{booktabs}
\usepackage[colorlinks,linkcolor=blue,citecolor=blue,urlcolor=blue]{hyperref}

\begin{document}
\allowdisplaybreaks
 \newcommand{\bq}{\begin{equation}}
 \newcommand{\eq}{\end{equation}}
 \newcommand{\bqn}{\begin{eqnarray}}
 \newcommand{\eqn}{\end{eqnarray}}
 \newcommand{\nb}{\nonumber}
 \newcommand{\lb}{\label}
 \newcommand{\f}{\frac}
 \newcommand{\p}{\partial}
\newcommand{\PRL}{Phys. Rev. Lett.}
\newcommand{\PLB}{Phys. Lett. B}
\newcommand{\PRD}{Phys. Rev. D}
\newcommand{\CQG}{Class. Quantum Grav.}
\newcommand{\JCAP}{J. Cosmol. Astropart. Phys.}
\newcommand{\JHEP}{J. High. Energy. Phys.}
\newcommand{\bea}{\begin{eqnarray}}
\newcommand{\ena}{\end{eqnarray}}
\newcommand{\beqa}{\begin{eqnarray}}
\newcommand{\eeqa}{\end{eqnarray}}
\newcommand{\red}{\textcolor{red}}

\newlength\scratchlength
\newcommand\s[2]{
  \settoheight\scratchlength{\mathstrut}%
  \scratchlength=\number\numexpr\number#1-1\relax\scratchlength
  \lower.5\scratchlength\hbox{\scalebox{1}[#1]{$#2$}}%
}

\title{Bounding the number of spacetime dimensions from precessing black hole binaries with the third-generation gravitational-wave detectors}

\author{Xu-Jie Zhu${}^{a, b}$}
\email{zhuxujie@zjut.edu.cn}

\author{Ji-Yu Song${}^{c}$}
\email{songjiyu@stumail.neu.edu.cn}

\author{Tao Zhu${}^{a, b}$}
\email{Corresponding author: zhut05@zjut.edu.cn}

\author{Xin Zhang${}^{c, d, e}$}
\email{zhangxin@neu.edu.cn}

\affiliation{${}^{a}$Institute for Theoretical Physics \& Cosmology, Zhejiang University of Technology, Hangzhou, 310023, China\\
${}^{b}$ United Center for Gravitational Wave Physics (UCGWP),  Zhejiang University of Technology, Hangzhou, 310023, China\\
 ${}^{c}$Liaoning Key Laboratory of Cosmology and Astrophysics, College of Sciences, Northeastern University, Shenyang 110819, China\\
${}^{d}$MOE Key Laboratory of Data Analytics and Optimization for Smart Industry, Northeastern University, Shenyang 110819, China \\
${}^{e}$National Frontiers Science Center for Industrial Intelligence, Northeastern University, Shenyang, 110819, China}

\date{\today}

\begin{abstract}

In the theories with extra dimensions, gravitational waves can leak into extra dimensions, resulting in a reduction in the amplitude of the observed gravitational waves. Such an effect modifies the standard luminosity distance of gravitational wave sources, which enables constraining extra dimensions by measuring both the luminosity distance and redshift from gravitational wave events. The main purpose of this paper is to assess the capacity of the third-generation gravitational wave detector network, Einstein Telescope and Cosmic Explorer, to constrain the theory of extra dimensions through observations of precessing binary black hole mergers. To this end, we generate a dataset of signals of precessing binary black hole mergers detectable by the detector network with one year of observations. We then employ the Fisher information matrix method for parameter estimation, together with a dark siren approach that obtains the redshift information from the galaxy catalog within a hierarchical Bayesian framework. Our results show that the precession in the binary system can significantly improve the precision of the luminosity distances and narrow their redshift ranges, thus leading to a tighter constraint on the extra dimensions. Based on this approach, we constrain the number of spacetime dimensions $D$ and the screening scale $R_{\rm c}$ of the extra dimensions, obtaining $D= 3.99^{+0.07}_{-0.06}$ and ${\rm log}_{\rm 10}R_c/{\rm Mpc} > 3.76$ at 68\% credible level. This result improves previous ones from an analysis with GWTC-3 by about one order of magnitude. This work provides theoretical foundations and projected sensitivities for future third-generation gravitational wave observations in exploring the effects of extra dimensions.

\end{abstract}

\maketitle

\section{Introduction}
\renewcommand{\theequation}{1.\arabic{equation}} \setcounter{equation}{0}

General relativity (GR) stands as the preeminent theory of gravity, having undergone rigorous experimental and observational validation across a diverse range of scales with remarkable precision \cite{Berti:2015itd, Wex:2014nva, Kramer:2016kwa, will2018}. Despite its empirical success, GR encounters significant challenges related to theoretical singularities and issues of quantization, as well as unresolved phenomena of dark matter and dark energy \cite{Bojowald:2007ky, Arun:2017uaw}. The development of new observational probes is therefore crucial for advancing our understanding of gravity and for testing possible extensions of GR. In this regard, the detection of gravitational waves (GWs) has opened up a completely new window. Based on the GW events observed to date, gravity can now be tested in ways independent of traditional electromagnetic probes, thereby enabling the formulation of a range of new theoretical frameworks aimed at confronting these challenges \cite{Cognola:2006eg, Copeland:2006wr, Frieman:2008sn}. 

The existence of GWs was first indirectly confirmed in the 1970s through the precise timing of binary pulsars \cite{Hulse:1974eb}. In 2015, the LIGO and Virgo detectors made the first direct observations of GWs from a binary black hole merger \cite{LIGOScientific:2016aoc}. To date, the LIGO-Virgo-KAGRA (LVK) collaboration has reported the detection of about 300 GW events \cite{LIGOScientific:2025slb}. Future ground-based third-generation detectors, such as the Einstein Telescope (ET) \cite{Evans:2021gyd} and Cosmic Explorer (CE) \cite{Branchesi:2023mws}, and space-based detectors, such as LISA \cite{Robson:2018ifk, LISACosmologyWorkingGroup:2022jok}, Taiji \cite{Ruan:2018tsw, Hu:2017mde}, and TianQin \cite{Liu:2020eko, TianQin:2020hid, Luo:2025ewp}, will extend the detection horizon considerably. GWs propagate through the universe largely unaffected by matter, carrying clean information about compact binary systems and providing a direct probe of strong-field gravity and modified theories of gravity.

The growing catalog of GW events has made GW astronomy a promising tool for testing theories of gravity with extra spacetime dimensions. In such theories, GWs can leak into the additional spacetime dimensions, altering the evolution of the amplitude of GWs and leading to a reduction in the amplitude of the observed GWs \cite{Deffayet:2007kf, Nishizawa:2017nef, Yu:2019jlb}. This effect arises in several well-motivated brane-world scenarios, including the Randall–Sundrum II model \cite{Randall:1999vf}, the Dvali–Gabadadze–Porrati model \cite{Dvali:2000hr}, and the Arkani-Hamed–Dimopoulos–Dvali model \cite{Arkani-Hamed:1998jmv}, etc. Although these theories are based on different physical assumptions and applicable scales, they all predict measurable deviations from four-dimensional GR. Such deviations in GWs can appear in terms of the amplitude-distance relation \cite{Deffayet:2007kf, Nishizawa:2017nef, Yu:2019jlb}, in the phase evolution of the waveform \cite{Nishizawa:2017nef, Du:2020rlx}, and in the damping structures of the ringdown signals \cite{Cardoso:2002pa, Chakraborty:2017qve, Mishra:2021waw, Arbelaez:2025gwj}. 

The gravitational leakage effect due to the existence of the extra spacetime dimension modifies the standard luminosity distance of GW sources, which enables constraining extra dimensions by measuring both the luminosity distance and redshift from GW events. 
Constraints on extra dimensions parameters have been extensively explored using the currently available LVK data, such as the characteristic length scale associated with gravitational leakage \cite{McWilliams:2009ym, Visinelli:2017bny} and the dimensions of spacetime \cite{Pardo:2018a, LIGO:2019f, MaganHernandez:2022, Corman:2020pyr, Corman:2021avn}. Note that the modified GW friction can also induce a modified luminosity distance of GWs, and the corresponding constraints were explored with LVK data or future detections, see refs.~\cite{Belgacem:2018lbp, Lagos:2019kds, Mastrogiovanni:2020mvm, Finke:2021aom, Chen:2023wpj, Mancarella:2022cgn, Leyde:2022orh, Ezquiaga:2021ayr, Mancarella:2021ecn, Chen:2024xkv, Liu:2023onj, Yang:2021qge, Narola:2023viz, Matos:2022uew, Lin:2024pkr, Zhu:2023rrx, Chen:2024pln, Zhang:2025kcw, Bian:2025ifp, Zhang:2024rel} and references therein for examples. The advent of next-generation detectors, including CE, ET, LISA, Taiji, and Tianqin, is expected to enhance the precision of these constraints. 

To constrain theories with extra dimensions, independent measurements of both the luminosity distance and the redshift of GW events are required. The luminosity distance can be inferred directly from the GW waveform \cite{Schutz:1986gp}. This method is called the standard siren, which can be used to measure cosmological parameters (the Hubble constant and dark energy parameters) \cite{Jin:2025dvf, Wang:2018lun, Zhang:2019ylr, Wang:2019tto, Zhang:2019loq, Zhao:2019gyk, Jin:2020hmc, Wang:2021srv}. But the precision of this method is limited by the distance–inclination degeneracy \cite{Bulla:2022ppy}. While electromagnetic counterparts to GW events can break this degeneracy \cite{Finstad:2018wid, LIGOScientific:2018hze}, most events are expected to remain without such counterparts, necessitating alternative approaches. Notably, spin precession in binary black hole systems has been shown to effectively mitigate the distance–inclination degeneracy, thereby improving distance determination \cite{Yun:2023, Raymond:2008im, Green:2020ptm}. In addition, when electromagnetic redshift measurements are unavailable, redshift information may be obtained statistically via methods such as spectral sirens \cite{Ezquiaga:2022zkx,  MaganaHernandez:2024uty} or by cross-matching with potential host galaxies \cite{DES:2020nay,  Gair:2022zsa, Palmese:2021mjm}, this relying on galaxy catalogs localization is known as the dark siren
\cite{Song:2022siz, Jin:2023sfc, Song:2025ddm, Song:2025bio}. Combining precise distance estimates from precessing systems with statistically derived redshift constraints thus provides a viable way to constrain extra-dimensional parameters even in the absence of electromagnetic counterparts.

In this paper, we explore the forecast of constraining the extra dimension parameters by using the precessing black hole binaries detected by the third-generation GW detectors (ET and CE). We simulated a dataset of GW signals of precessing black hole mergers detectable by the detector network with one year of observation. In particular, we perform an analysis of these GW signals based on the Fisher information matrix (FIM) \cite{Poisson:1995, Finn:1993, Cutler:1994} for estimation of source parameters. Since source redshift cannot be directly obtained in the absence of electromagnetic counterparts, we adopt the dark siren approach, which statistically combines GW distance measurements with potential host galaxy redshift distributions from galaxy catalogs within a hierarchical Bayesian inference scheme \cite{Finke:2021aom, Chen:2023wpj, Mancarella:2022cgn, Mancarella:2022cgn}. In addition, waveform models play an important role in the accuracy of parameter estimation. We consider the GW waveform model that includes spin precession effects to improve the accuracy of the measurement of the luminosity distance.

This paper is organized as follows. In Sec.~II, we provide a brief overview of GW propagations with the presence of the extra dimensions. In Sec.~\ref{sec3}, we describe the application of the FIM for parameter estimation from simulated GW signals. Sec.~\ref{sec4} presents the dark siren method within a hierarchical Bayesian framework, which we use to infer source redshifts without electromagnetic counterparts and to obtain constraints on extra-dimensional parameters. The resulting constraints are presented and discussed in Sec.~\ref{sec5}. Finally, conclusions and a summary of this work are given in Sec.~ \ref{sec6}.

Throughout this paper, we adopt the cosmological parameters from the Planck 2018 results \cite{Planck:2018vyg}, $H_0 = 67.27\;\text{km}\,\text{s}^{-1}\,\text{Mpc}^{-1}$, $\Omega_m = 0.3166$, and $\Omega_k = 0$.

\section{Extra dimension in the GW propagation \label{sec2}}
\renewcommand{\theequation}{2.\arabic{equation}} \setcounter{equation}{0}

In this section, we briefly review how the presence of extra dimensions modifies the propagation of GWs, specifically by altering their amplitude evolution, which in turn affects the observed GW luminosity distance.

In GR, the amplitude of GWs scales inversely with luminosity distance, so that the strain $h$ of GWs is related to the source's luminosity distance $d^{\rm GW}_L$ by,
\begin{equation} 
     h_{\rm GR} \propto \frac{1}{d^{\rm GW}_{\rm L}}.
\end{equation}
In a higher-dimensional spacetime, however, GWs can leak into extra dimensions. Flux conservation implies that their amplitude decays with distance according to a modified power law. Phenomenologically, this effect can be described by the relation \cite{Corman:2021avn},
\begin{equation}
    d^{\rm GW}_{\rm L} \propto d^{\rm EM}_{\rm L} \left( \frac{d^{\rm EM}_{L}}{(1+z)} \right)^{(D-4)/2},
\label{dl_em=dl_gw}    
\end{equation}
where $D$ denotes the number of spacetime dimensions, $z$ is the redshift of the GW source, and $d^{\rm EM}_{\rm L}$ denotes the standard luminosity distance to the source in a flat $\Lambda$CDM cosmology, which is given by, 
\begin{equation}
d^{\rm EM}_{\rm L} = \frac{1+z}{H_0} \int_0^z \frac{dz'}{\sqrt{\Omega_\Lambda + \Omega_m (1+z')^3}}.
\label{dl_em}
\end{equation}
In the four dimensional limit $D = 4$, the usual relation $d^{\text{GW}}_{\rm L} = d^{\rm EM}_{\rm L}$ is recovered.

In the Solar System, GR remains consistent with precision tests, placing strong constraints on possible higher-dimensional gravitational effects \cite{DiValentino:2021izs, Overduin:2000gr}. To reconcile higher-dimensional theories with these local bounds, a characteristic transition scale $R_{c}$ is introduced. At distances much smaller than $R_{c}$, spacetime effectively appears four-dimensional, and the GW luminosity distance $d^{\rm GW}_L$ coincides with its electromagnetic counterpart $d^{\rm EM}_L$. To describe a smooth transition between the four-dimensional regime at small scales and the higher-dimensional regime at large scales, we follow the phenomenological model of Refs. \cite{Pardo:2018a, LIGO:2019f, Deffayet:2007kf} and generalize Eq.~(\ref{dl_em=dl_gw}) to
\begin{equation}
d^{\rm GW}_{\rm L} = d^{\rm EM}_{\rm L} \left[ 1 + \left( \frac{d^{\rm EM}_{\rm L}}{R_c (1 + z)} \right)^n \right]^{(D-4)/(2n)} .
\label{dl_gw_dl_em}
\end{equation}
Here, $R_c$ sets the characteristic distance beyond which extra-dimensional effects become significant, and the index $n$ controls the sharpness of the transition between the four- and higher-dimensional behaviors.

Several constraints on the extra-dimensional parameters $(D, R_c)$ have been derived from analyses of GW events with electromagnetic counterparts \cite{Pardo:2018a, LIGO:2019f}, as well as from statistical methods such as the spectral siren approach \cite{MaganHernandez:2022, Leyde:2022orh, Chen:2023wpj}. Forecasts for improving these constraints with future GW detectors have also been presented in earlier studies (see, e.g., Refs.~\cite{Corman:2020pyr, Corman:2021avn}). A summary of the corresponding results is presented in Table~\ref{results}.

In this work, we adopt the phenomenological relation given by Eq.~\eqref{dl_gw_dl_em} and, for analytical clarity, set the transition exponent $n = 1$. Our aim is to investigate how future third-generation GW observations can further refine the constraints on the effects of extra dimensions.

\begin{table*}
\caption{Summary of previous constraints on the extra-dimensional parameters $D$ and $R_c$ obtained from GW observations using different datasets and methods, such as electromagnetic counterparts and spectral siren methods. Dashes indicate parameters not constrained in the respective analysis.}
\label{results}
\centering
\begin{ruledtabular}
\begin{tabular}{lcc}
Methods for determining redshifts & $D$ & $R_c$ \\
\hline\\[-6pt]
\multirow{3}{*}{Electromagnetic counterparts \cite{Pardo:2018a}} 
 & $D=4.02^{+0.07}_{-0.10}$ (SHoES) & — \\[2pt]
 & $D=3.98^{+0.07}_{-0.09}$ (Planck) & — \\[2pt]
 & — & {$R_c\gtrsim 20\ \mathrm{Mpc}$}\\[2pt]
\hline\\[-6pt]
\multirow{2}{*}{spectral siren method \cite{MaganHernandez:2022}} 
 & $D = 3.95^{+0.09}_{-0.07}$ & independent of $R_c$ \\[2pt]
 & $D=4.23^{+1.50}_{-0.57}$ 
 & $\log_{10}(R_c/\mathrm{Mpc}) = 4.14^{+0.55}_{-0.86}$ with $\log_{10} n = 0.86^{+0.73}_{-0.84}$\\[2pt]
 \hline\\[-6pt]
 spectral siren method \cite{Leyde:2022orh}
 & $D = 4.6^{+2.6}_{-0.8}$ & — \\[2pt]
 \hline\\[-6pt]
 spectral siren method \cite{Chen:2023wpj}
 & $D = 4.07^{+1.01}_{-0.23}$ 
 & $\log_{10}(R_c/\mathrm{Mpc}) = 3.93^{+1.55}_{-0.55}$ \\[2pt]
\end{tabular}
\end{ruledtabular}
\end{table*}

\section{Parameter estimations with simulated GW singals} \lb{sec3}
\renewcommand{\theequation}{3.\arabic{equation}} \setcounter{equation}{0}

\subsection{Simulated signals from precessing black hole binaries and detector network}
 
So far, the LVK collaboration has reported the observation of approximately 300 GW events \cite{LIGOScientific:2025slb}, all originating from the mergers of compact binary systems. While only a small fraction of these events may exhibit spin-precession effects \cite{Hannam:2021, Islam:2023}, such effects, when present, can improve the measurement accuracy of certain physical parameters in GW observations, including luminosity distance \cite{Yun:2023}. To systematically study the constraining power of GW events with precession effects on extra-dimensional parameters, we simulate binary black hole signals using the phenomenological waveform model IMRPhenomPv3 \cite{Khan:2019}, which builds on previous models \cite{Hannam:2014, Santamaria:2010, Husa:2016I, Khan:2016II} and integrates a double-spin approach to add the latest understanding of precession dynamics. We generate populations with both high and low precession to compare their respective impacts on parameter estimation and on constraints of extra-dimensional parameters.

For our detector network, we consider a third-generation ground-based network with two detectors, ET and CE, to constrain extra dimensions. The performance parameters of these detectors are summarized in Table \ref{detectors}, and further details on ET and CE can be found in reference \cite{Muttoni:2023}. For CE, we adopt a 40 km arm length configuration to achieve higher detection accuracy. In terms of sensitivity curves, CE uses the CE2 (silicon) sensitivity curve \cite{Abbott:2017eg}, predicted by \textsc{gwinc}, while ET employs the ET-D curve \cite{Punturo:2010eg} to characterize strain sensitivity across the frequency spectrum. Waveforms are generated using \textsc{Bilby} \cite{Ashton:2018jfp}, and parameter estimation is performed via the Fisher information matrix (FIM) using the respective power spectral densities (PSD) of each detector.

\begin{table*}
\caption{Key performance parameters of the GW detectors (ET and CE) used in our analysis.}
\label{detectors}
\centering  
\begin{ruledtabular}
\begin{tabular}{lccccccc}
Detector & Abbreviation & Latitude & Longitude & Arm length & x-arm azimuth & y-arm azimuth & Sensitivity \\
\hline\\[-6pt]
Einstein Telescope & ET & {43.63} & {10.50} & {10}{km} & {70.57} & {130.57} & ET-D \\[2pt]
Cosmic Explorer & CE & {46.46} & {-119.41} & {40}{km} & {126.00} & {215.994} & CE2 \\[2pt]
\end{tabular}
\end{ruledtabular}
\end{table*}

We simulate one year of detectable precessing binary black hole events for the ET+CE network. The source parameters of these events, like masses, spins, redshifts, etc., are generated by the prescribed population models. The masses sampled from the Power Law + Peak mass model\footnote{Our simulate is based on the median results constrained by the current population model.} \cite{LIGOScientific:2018jsj}, the spin magnitudes and orientations follow the Default spin model \cite{Talbot:2017yur}, the redshifts are generated according to a Power Law redshift distribution \cite{Stasenko:2024pzd}, and the luminosity distances are obtained using Eq. (\ref{dl_em}). Each event is characterized by 15 source parameters: $\{m_1, m_2, d_L, \iota, \text{ra}, \text{dec}, \psi, a_1, a_2, \theta_1, \theta_2, \phi_{\text{JL}}, \phi_{12}, t_c, \phi_c\}$. Here, $m_1$ and $m_2$ are the component masses, $d_L$ the luminosity distance, $\text{ra}$ and $\text{dec}$ the sky location, $\psi$ the polarization angle, $\iota$ the inclination angle of the binary system, $a_1$ and $a_2$ the dimensionless spin magnitudes, $\theta_1$ and $\theta_2$ the angles between the two spin vectors and the orbital angular momentum, $\phi_{\text{JL}}$ the misalignment angle between the two spin vectors, $\phi_{12}$ the precession phase between the two spins, and $t_c$ and $\phi_c$ the merger time and phase. 

The effective precessing spin of the binary system is related to the parameters $\theta_1$, $\theta_2$, $a_1$, $a_2$, and $q = m_1/m_2$ \cite{Hannam:2014, Schmidt:2015eg}
\begin{equation}   
\chi_p \equiv \max \left\{ a_1 \sin\theta_1,\  \frac{q(4q + 3)}{4 + 3q} a_2 \sin\theta_2 \right\}.
\end{equation}
In our simulations, $\theta_1$, $\theta_2$, $a_1$, and $a_2$ are chosen such that $\chi_p$ is uniformly distributed in $(0,1)$. The remaining angular parameters ${\theta_{\text{JN}}, \text{ra}, \text{dec}, \psi, \phi_{\text{JL}}, \phi_{12}}$ as well as $t_c$ and $\phi_c$ are randomly sampled over their physical ranges to ensure a representative and diverse dataset.

\subsection{Estimations of luminosity distances}

In this section, we introduce the FIM method used for parameter estimation, following the approach established for GWs from compact binary systems in Ref.~\cite{Poisson:1995, Finn:1993, Cutler:1994}. The components of FIM are defined as the noise-weighted inner product between the partial derivatives of the GWs waveform $h$ with respect to source parameters $\theta_i$,
\begin{equation}
    \Gamma_{ij} = \left( \frac{\partial h}{\partial \theta_i} \middle| \frac{\partial h}{\partial \theta_j} \right),
    \label{FIM}
\end{equation}
where the inner product is weighted by the PSD of the detector noise. The inverse of the FIM, $\Gamma^{-1}$, provides an approximation to the covariance matrix of the parameter estimates. The diagonal elements of $\Gamma^{-1}$ correspond to the variances of the respective parameters, which quantify their statistical uncertainties.

First, we define the noise-weighted inner product between two GW signals, $h_1$ and $h_2$, 
\begin{equation}
    (h_1|h_2) = 2 \int_{f_{\min}}^{f_{\max}} \frac{\tilde{h}^*_1(f)\tilde{h}_2(f) + \tilde{h}^*_2(f)\tilde{h}_1(f)}{S_n(f)}  df,
    \label{eq:inner_product}
\end{equation}
where $\tilde{h}_1(f)$ and $\tilde{h}_2(f)$ are the frequency-domain signals obtained by applying the Fourier transformation to the time-domain GWs signal $h(t)$, $S_n(f)$ is the PSD of the detector, and the $^*$ denotes the complex conjugation. The integration limits, $f_{\max}$ and $f_{\min}$, represent the maximum and minimum threshold frequencies of the detector, respectively. The signal-to-noise ratio (SNR) for a given signal $h$ is then given by
\begin{equation}
    \rho = (h|h)^{1/2}.
\end{equation}

The components of the FIM are evaluated using Eq.~(\ref{FIM}),  where the parameters $\theta_i$ and $\theta_j$ belong to the set of GW waveform parameters $\{m_1, m_2, d_L, \iota, \text{ra}, \text{dec}, \psi, t_c, \phi_c\}$. Since we are using numerical waveforms, the forward difference method is employed to compute the derivatives of the corresponding parameters
\begin{equation}
    \frac{\partial {h}}{\partial \theta_i} \approx \frac{{h}(\theta_i + \Delta) - {h}(\theta_i)}{\Delta}
\end{equation}
where $\Delta$ is a small quantity. Then, the standard deviation $\Delta\theta_i$ (i.e., the $1\sigma$ uncertainty) for parameter $\theta_i$ can be calculated, which is obtained by taking the square root of the corresponding diagonal elements of the inverse FIM
\begin{equation}
    \Delta\theta_i = \sqrt{(\Gamma^{-1})_{ii}},
\end{equation}
where $\Gamma^-1$ is the inverse of FIM. In Eq. (\ref{eq:inner_product}), the lower frequency ($f_{\text{min}}$) is set to the detector’s minimum sensitivity threshold, while the upper frequency ($f_{\text{max}}$) is set to 2048 Hz.

\begin{figure}[htbp]
  \centering
  \includegraphics[width=\linewidth]{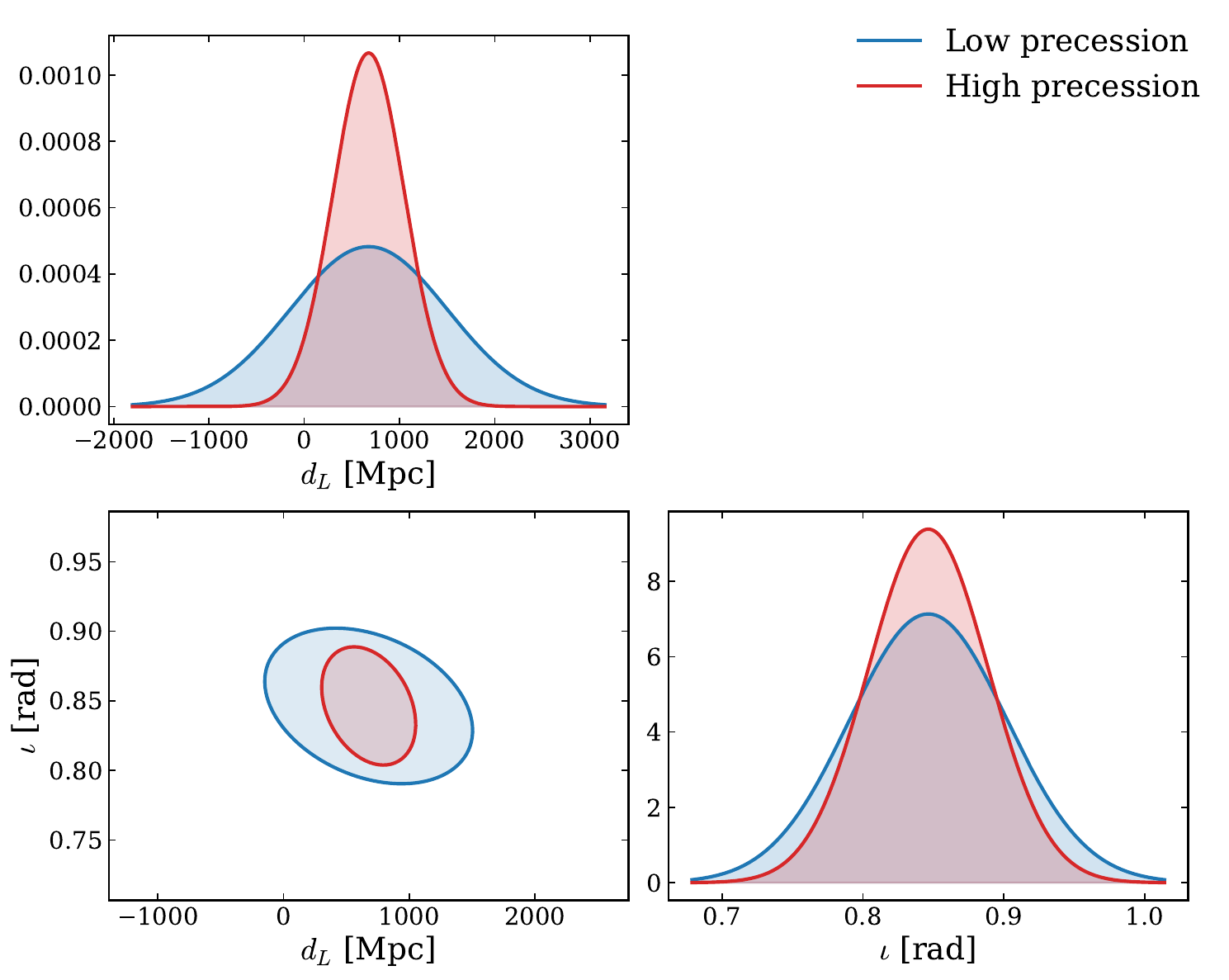}
  \caption{Corner plot showing the posterior distributions of the luminosity distance $d_L$ and inclination angle $\iota$ obtained from FIM analysis. The blue and red shaded regions correspond to the $1\sigma$ credible intervals for a low-precession event ($\chi_p = 0.01$) and a high-precession event ($\chi_p \sim 0.7$), respectively. The improved precision for the high-precession case illustrates how precession helps break the distance–inclination degeneracy.}
  \label{fim}
\end{figure}

The main purpose of the FIM analysis is to estimate the measurement uncertainties of the luminosity distance, right ascension, and declination of GW sources. To study how precession improves the precision of luminosity distance estimation, we select binary black hole events with an SNR $\rho > 100$ and effective precession spin $\chi_p$ in the range of $[0.3, 0.8]$. For comparison, we also analyzed a set of similar events with low precessing spin ($\chi_p = 0.01$). The parameter estimation results for both a typical low- and high-precession events are shown in Fig.~\ref{fim}. As illustrated, incorporating precession effects helps break the degeneracy between luminosity distance and inclination angle, leading to improved precision in the distance measurement. Consequently, high-precession events are expected to yield tighter constraints than their low-precession counterparts.

\section{The dark siren method for infering redshift distributions}\lb{sec4}
\renewcommand{\theequation}{4.\arabic{equation}} \setcounter{equation}{0}

In this work, we employ the dark siren method to infer redshifts of GW sources that lack electromagnetic counterparts. Since GW observations typically yield broad sky localization regions, the host galaxy of a given source can not be uniquely identified. By cross-matching the GW localization area with a galaxy catalog, we obtain a set of candidate host galaxies, which collectively provide a redshift prior for each GW source. Combining this prior with the GW luminosity distance measurement, one is able to constrain both cosmological and modified gravity parameters within a hierarchical Bayesian framework. The framework adopted here follows Ref.~\cite{chen:2017rfc}.

According to Bayes’ theorem, the posterior distribution of cosmological or modified gravity parameter, denoted by ${\boldsymbol \Lambda}$, is given by
\begin{equation}
\begin{aligned}
    p(\boldsymbol{\Lambda}|\{X_{\rm GW}\})&\propto p_0(\boldsymbol{\Lambda})p(\{X_{\rm GW}\}|\boldsymbol{\Lambda})\\
    &\propto p_0(\boldsymbol{\Lambda})\prod_i^{N_{\rm GW}}p(X_{{\rm GW},i}|\boldsymbol{\Lambda}),
\end{aligned}
\end{equation}
where $p_0(\boldsymbol{\Lambda})$ is the prior on ${\boldsymbol \Lambda}$, $N_{\rm GW}$ is the number of GW events in the GW data set $\{X_{\rm GW}\}$, and $p(X_{{\rm GW}, i}|\boldsymbol{\Lambda})$ is the likelihood for the $i$-th event. The likelihood function of a single GW event is given by
\begin{equation}
\begin{aligned}
    \mathcal{L}(X_{\rm GW}|\boldsymbol{\Lambda})=&\frac{1}{\beta(\boldsymbol{\Lambda})}\int {\rm d}d_L{\rm d}z{\rm d}\alpha{\rm d}\delta \Big[p(X_{\rm GW}|d_L^{\rm obs},\alpha,\delta)\\
    & \;\;\;\; \times  p_0(z,\alpha,\delta)\delta \big(d_L^{\rm obs}-d_L^{\rm thro}(z,\boldsymbol{\Lambda}) \big) \Big],
\end{aligned}
\end{equation}
where $p(X_{\rm GW}|d_L^{\rm obs},\alpha,\delta)$ is the marginalized likelihood of the GW data, and the Dirac delta function $\delta\left[d_L^{\rm obs}-d_L^{\rm theo}(z,\boldsymbol{\Lambda})\right]$ relates the observed luminosity distance to its theoretical prediction for a given $(z,\boldsymbol{\Lambda})$. The prior $p_0(z,\alpha,\delta)$ on the source redshift, right ascension, and declination, is constructed as
\begin{equation}
\begin{aligned}
    p_0(z,\alpha,\delta) = & f_{\rm com}p_{\rm cat}(z,\alpha,\delta) \\
    &+(1-f_{\rm com})p_{\rm miss}(z,\alpha,\delta|\boldsymbol{\Lambda}).
    \end{aligned}
\end{equation}
Here $p_{\rm cat}(z,\alpha,\delta)$ is the redshift prior provided by the galaxy catalog, which is expressed as
\begin{equation}
    p_{\rm cat}(z,\alpha,\delta) = \frac{1}{N_{\rm in}}\sum_{j=1}^{N_{\rm in}}\mathcal{G}(\hat{z}_j,\sigma_{z,j})\delta(\alpha-\hat{\alpha}_j)\delta(\delta-\hat{\delta}_j),
\end{equation}
where $N_{\rm in}$ is the number of candidate host galaxies within the GW localization region. The Gaussian $\mathcal{G}(\hat{z}_j,\sigma{z,j})$ models the redshift uncertainty of the $j$-th galaxy. We neglect galaxy sky-localization uncertainties, following Ref.~\cite{chen:2017rfc}. The term $p_{\rm miss}(z,\alpha,\delta\mid\boldsymbol{\Lambda})$ describes the redshift distribution of galaxies missing from the catalog and is given by
\begin{equation}
    p_{\rm miss}(z|\boldsymbol{\Lambda})\propto[1-p_{\rm com}(z,\boldsymbol{\Lambda})]\frac{{\rm d}V_{\rm c}}{{\rm d}z{\rm d}\Omega},
\end{equation}
where $p_{\rm com}(z)$ denotes the redshift-dependent completeness of the galaxy catalog. Owing to the finite sensitivity of survey telescopes, distant and faint galaxies are increasingly missed, causing the catalog completeness to decrease with redshift. We estimate $p_{\rm com}(z)$ using a Schechter luminosity function \cite{Schechter:1976iz} with slope $\alpha=-1.07$ and characteristic luminosity $L_* = 1.2\times10^{12}\ h^{-2}L_{\odot}$, where $L_{\odot}$ is the solar luminosity. We impose a lower luminosity cutoff of $0.001, L_*$ and adopt an apparent-magnitude detection threshold of $m_{\rm th}=26.3$, corresponding approximately to the $5\sigma$ depth of next-generation surveys such as the China Space Station Survey Telescope \cite{CSST:2025ssq}.

The term $f_{\rm com}$ is a completeness fraction given by
\begin{equation}
    \begin{aligned}
        f_{\rm com}=\frac{4\pi}{V_{\rm c}^{\rm tot}}\int_{z_{\rm min}}^{z_{\rm max}}p_{\rm com}(z,\boldsymbol{\Lambda}){\rm d}z\frac{{\rm d}V_{\rm c}}{{\rm d}z},
    \end{aligned}
\end{equation}
where $z_{\rm min}$ and $z_{\rm max}$ is the minimum and maximum redshift of all the candidate host galaxies of a GW source, and $V_{\rm c}^{\rm tot}$ is the total comoving volume enclosed with $[z_{\rm min},z_{\rm max}]$.

The factor $\beta(\boldsymbol{\Lambda})$ corrects for the selection effects of GW observations and is given by
\begin{equation}
    \beta(\boldsymbol{\Lambda}) = \int p_{\rm det}^{\rm GW}(d_L)\delta(d_L-d_L(z,\boldsymbol{\Lambda})){\rm d}d_L{\rm d}z,
\end{equation}
where $p_{\rm det}^{\rm GW}(d_L)$ denotes the detection probability of GW events at luminosity distance $d_L$. We evaluate $p_{\rm det}^{\rm GW}(d_L)$ using Monte Carlo integration following Ref.~\cite{Gray:2019ksv}.

\begin{table}
\centering
\caption{Prior configurations adopted in the hierarchical Bayesian inference. The wide prior treats $H_0$ as a free parameter with a uniform distribution over $[60,90]\,\mathrm{km\,s^{-1}\,Mpc^{-1}}$, while the fixed prior sets $H_0$ to its fiducial value. The matter density $\Omega_m$ is fixed in both cases, and uniform priors are assumed for $D$ and $R_c$ over the ranges shown.}
\label{priors}
\begin{ruledtabular}
    \begin{tabular}{lcc}
Parameters & Wide prior & Fixed prior \\
\hline\\[-6pt]
$H_0$ & [60, 90] & 67.27 \\[2pt]
$\Omega_m$ & 0.3166 & 0.3166 \\[2pt]
$D$ & [3, 11] & [3, 11] \\[2pt]
$R_c$ & [20, 10000] & [20, 10000] \\[2pt]
\end{tabular}
\end{ruledtabular}
\end{table}


\section{RESULTS AND DISCUSSION}\lb{sec5}
\renewcommand{\theequation}{5.\arabic{equation}} \setcounter{equation}{0}

We employ the FIM to estimate the uncertainties in luminosity distance, right ascension, and declination from simulated GW signals, while independently identifying their potential host galaxies to obtain their redshift distributions. By combining the redshift and luminosity distance measurements, we perform hierarchical Bayesian inference using \textbf{emcee} \cite{ForemanMackey:2013} to constrain the extra-dimensional parameters $(D, R_c)$. Two different prior configurations (specifying as wide and fixed $H_0$ priors) are considered, as summarized in Table~\ref{priors}. For each prior setup, we analyze three distinct GW event samples: low-precession, high-precession, and a combined set, with the results discussed in separate subsections below.

\subsection{Results with wide $H_0$ prior}

\begin{figure*}[htbp]
  \centering
  \includegraphics[width=\linewidth]{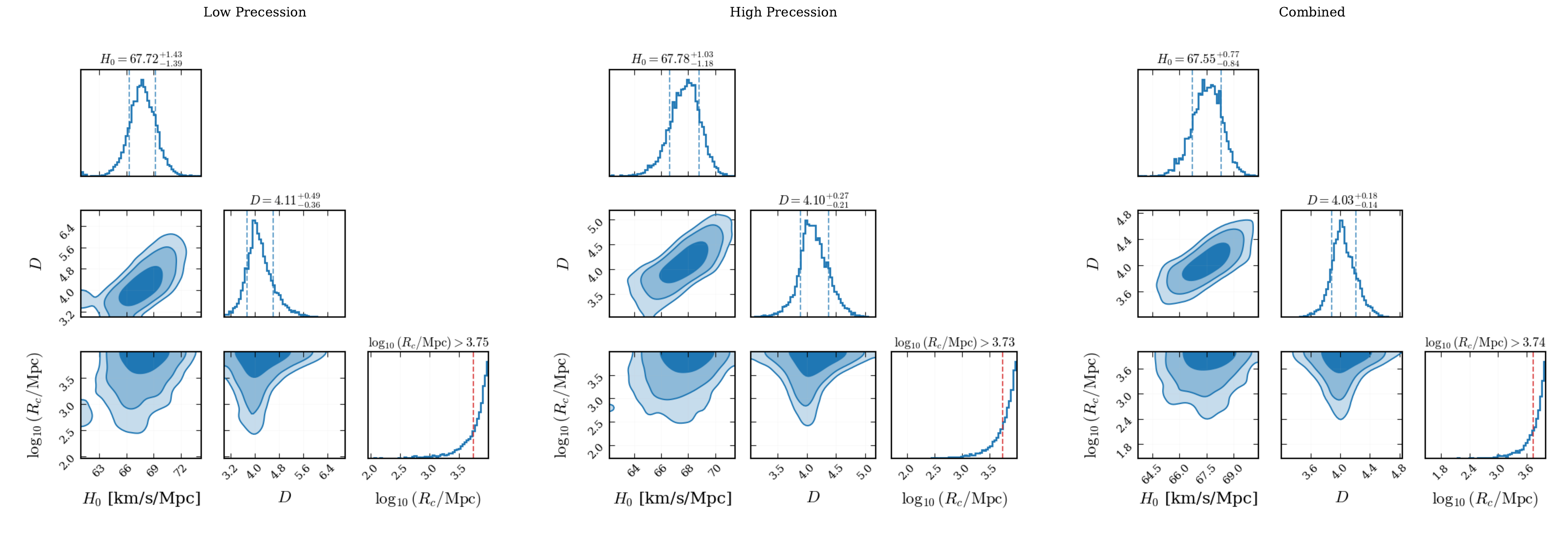}
 \caption{Joint posterior constraints on the parameters $(H_0, D, R_c)$ obtained from the combined GW sample under wide priors. The blue dashed lines indicate the $68\%$ credible intervals for $H_0$ and $D$, while the red dashed line marks the $68\%$ lower bound on $R_c$.}
  \label{wide_p}
\end{figure*}

We first present the constraints obtained under wide $H_0$ priors, which include the Hubble parameter $H_0$ as a free parameter in the joint inference. Figure~\ref{wide_p} summarizes the corresponding posterior distributions for the three parameters, while Table~\ref{wide_fixed} summarizes the corresponding intervals at the 68\% credible level for the three event samples.

\begin{table*}
\centering
\caption{Summary of constraints on the extra-dimensional parameters $D$ and $\log_{10}(R_c/\mathrm{Mpc})$ under both wide and fixed $H_0$ priors, for the low-precession, high-precession, and combined event samples. The $H_0$ constraints under the wide prior are also shown. All uncertainties and lower bounds correspond to the $68\%$ credible level.}
\label{wide_fixed}
\begin{ruledtabular}
\begin{tabular}{lccccc}
\multicolumn{1}{c}{\multirow{2}{*}{Event Type}} & \multicolumn{3}{c}{\textbf{Wide $H_0$ Priors}} & \multicolumn{2}{c}{\textbf{Fixed $H_0$ Priors}} \\
\cline{2-4} \cline{5-6}\\[-6pt]
\multicolumn{1}{c}{} & $D$ & $\log_{10}(R_c/\mathrm{Mpc})$ & $H_0$ & $D$ & $\log_{10}(R_c/\mathrm{Mpc})$ \\[2pt]
\hline\\[-6pt]
Low Precessing  & $4.11^{+0.49}_{-0.36}$ & $> 3.75$ & $67.72^{+1.43}_{-1.39}$ & $4.00^{+0.20}_{-0.17}$ & $> 3.74$ \\[2pt]
High Precessing & $4.10^{+0.27}_{-0.22}$ & $> 3.73$ & $67.78^{+1.03}_{-1.18}$ & $4.02^{+0.10}_{-0.09}$ & $> 3.75$ \\[2pt]
Combined        & $4.03^{+0.18}_{-0.14}$ & $> 3.74$ & $67.55^{+0.77}_{-0.84}$ & $3.99^{+0.07}_{-0.06}$ & $> 3.76$ \\[2pt]
\end{tabular}
\end{ruledtabular}
\end{table*}

With the wide prior, we jointly constrain $H_0$, $D$, and $R_{\rm c}$. We first consider the low-precession sample, which contains 657 GW events with $\rho > 100$. Due to the lower precision in luminosity distance parameter estimation compared to the precessing system, the sample yields the weakest constraints among the three cases,
 \begin{equation}
    \begin{gathered}
        H_0 = 67.55 ^{+1.43}_{-1.39}, \\
        D = 4.10^{+0.49}_{-0.36}, \\
        \log_{10}(R_c) > 3.75.
    \end{gathered}
\end{equation}

For the high-precession sample, 1232 events satisfy the selection criteria. The improved distance precession due to the presence of the significant precessing effect leads to tighter constraints, 
\begin{equation}
    \begin{gathered}
        H_0 = 67.55 ^{+1.03}_{-1.18}, \\
        D = 4.10^{+0.27}_{-0.22}, \\
        \log_{10}(R_c) > 3.73,
    \end{gathered}
\end{equation}
Compared to the low-precession case, the uncertainties on $H_0$ and $D$ are reduced by factors of approximately 1.4 and 1.8, respectively, while the bound on $R_c$ remains comparable.

The tightest overall constraints are obtained by combining precessing and non-precessing events, which increases the statistical sample size while retaining the benefits of precession for parameter estimation. For this combined sample, we obtain,
\begin{equation}
    \begin{gathered}
        H_0 = 67.55 ^{+0.77}_{-0.84}, \\
        D = 4.03^{+0.18}_{-0.14}, \\
        \log_{10}(R_c) > 3.74,
    \end{gathered}
\end{equation}
Relative to the high-precession only sample, the uncertainties on $H_0$ and $D$ are further reduced by factors of 1.4 and 1.5, respectively. Compared to the analysis with different prior setup using GWTC-3 data\footnote{Although we have made a rough comparison with their results, their approach differs from ours in that they infer redshifts from the mass distribution features of the binary black hole population and simultaneously constrain \(D\), \(R_c\), and \(n\) within a Bayesian framework.} \cite{MaganHernandez:2022}, which reported 
\begin{equation}
D = 4.23^{+1.50}_{-0.57},
\end{equation}
the precision on the spacetime dimension $D$ from our analysis with the wide prior is improved by a factor of 6.5.

\subsection{Results with fixed $H_0$ prior}

\begin{figure*}[htbp]
  \centering
  \includegraphics[width=\linewidth]{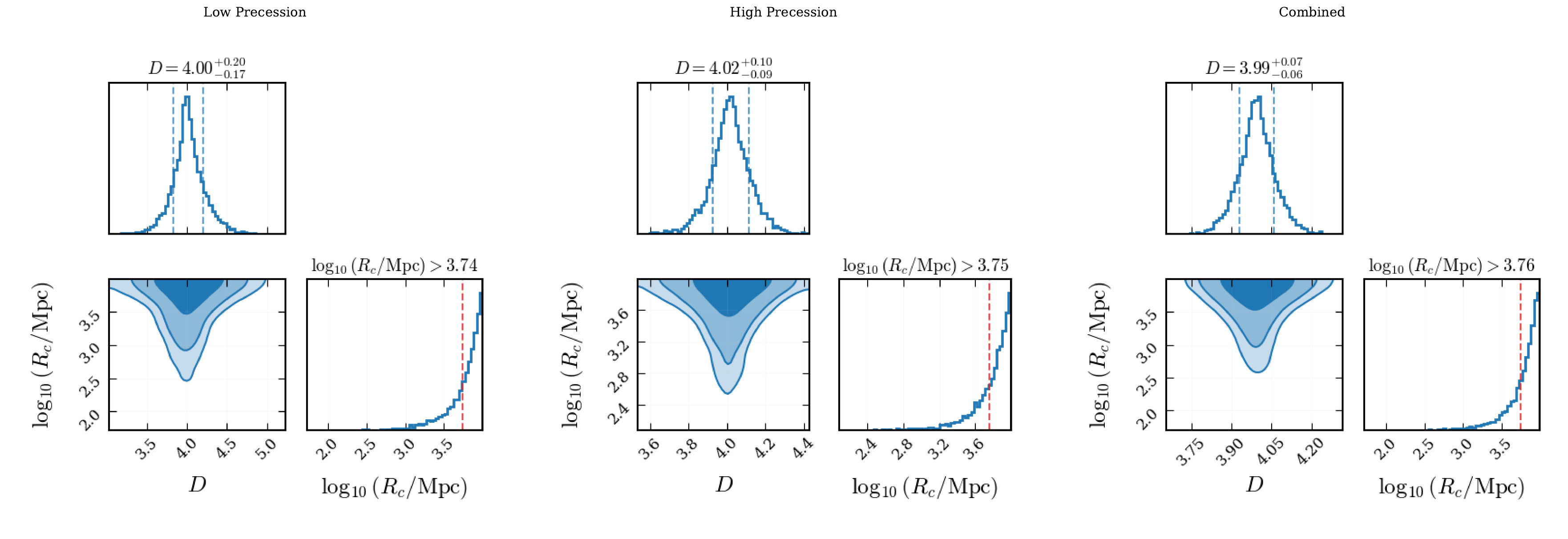}
\caption{Posterior distribution of the extra-dimensional parameters $(D, R_c)$ obtained from the combined GW sample under fixed $H_0$ prior. The blue dashed lines mark the $68\%$ credible interval for $D$, while the red dashed line indicates the $68\%$ lower bound on $R_c$.}
  \label{fix_p}
\end{figure*}

We now present the constraints obtained under a fixed prior on the Hubble constant, setting $H_0 = 67.27\,\mathrm{km\,s^{-1}\, Mpc^{-1}}$. Figure~\ref{fix_p} shows the posterior distributions for $(D, R_c)$, and Table~\ref{wide_fixed} summarizes the 68\% credible intervals.

We first consider the low-precession sample, which contains 657 events with $\rho > 100$. The resulting constraints are
\begin{equation}
    D = 4.00^{+0.20}_{-0.17}, \qquad
    \log_{10}(R_c/\mathrm{Mpc}) > 3.74 .
\end{equation}

For the high-precession sample (1232 events), the improved parameter estimation from precession yields tighter bounds:
\begin{equation}
    D = 4.02^{+0.10}_{-0.09}, \qquad
    \log_{10}(R_c/\mathrm{Mpc}) > 3.75 .
\end{equation}
Relative to the low-precession case, the uncertainty on $D$ is reduced by approximately a factor of two. 

The tightest constraints are again obtained by combining both samples, which benefits from the increased sample size while retaining the improved distance precision from precessing systems. For the combined sample, we find
\begin{equation}
    D = 3.99^{+0.07}_{-0.06}, \qquad
    \log_{10}(R_c/\mathrm{Mpc}) > 3.76 .
\end{equation}
This result slightly improves the constraints obtained from the precession events.  Compared to the wide-prior results, the uncertainty on $D$ improves by a factor of 2.5, while the lower bound on $R_c$ remains consistent across all cases. It also improves the previous result from an analysis with the GWTC-3 event by at least one order of magnitude. 

In summary, our analysis demonstrates that high-precession events enhance parameter estimation through improved spatial localization and distance precision, yielding constraints on the spacetime dimension $D$ that are approximately twice as tight as those from low-precession events. Combining both precessing and non-precessing samples further increases the statistical power, leading to the most stringent limits obtained in this work. All results remain compatible with ($D=4$) within the $68\%$ credible intervals. In addition, our tightest constraints are comparable in precision to those of GW170817 and its electromagnetic counterpart; future joint analyses with standard sirens are expected to get tighter constraints.

\section{CONCLUSION AND OUTLOOK}\lb{sec6}
\renewcommand{\theequation}{6.\arabic{equation}} \setcounter{equation}{0}

In this work, we investigate how extra dimensions affect the propagation of GWs and their potential constraints from future ground-based GW detectors. The effects of extra dimensions alter the damping rate of GW amplitudes, causing the GW luminosity distance to deviate from its counterpart in GR. By simultaneously measuring the luminosity distance and redshift of GW sources, one can test these modifications using GW events as standard sirens. Our main objective is to assess the constraining power of the next-generation ground‑based detector network, ET and CE, using observations of precessing binary black hole mergers.

We simulate one year of detectable binary black hole events for an ET+CE network. Using the FIM approach, we estimate the measurement uncertainties in luminosity distance and sky localization for each event. Based on the inferred localization regions, we cross‑match with a simulated galaxy catalog to obtain redshift distributions from potential host galaxies. Finally, we perform a hierarchical Bayesian analysis that combines the redshift information and luminosity‑distance measurements from precessing events meeting our selection criteria, thereby deriving constraints on extra‑dimensional parameters.

In the Bayesian inference, we adopt two different priors for the Hubble constant $H_0$, while fixing $\Omega_m = 0.3166$. We find that high‑precession events improve the constraints on the spacetime dimension $D$ by approximately a factor of two relative to low‑precession events. All posteriors for $D$ remain consistent with GR ($D=4$) within the $68\%$ credible intervals.

This study highlights the potential of third‑generation GW detector networks to probe extra‑dimensional theories, providing a theoretical foundation and projected sensitivities for future observations. In the present analysis, we have focused on waveform models that include spin‑precession effects to quantify their impact on parameter estimation. A natural extension for future work will be to incorporate higher‑order modes together with precession, as previous studies indicate that detectable higher‑order modes can further break the distance–inclination degeneracy and thus improve the precision of luminosity‑distance measurements. Combining precession and higher‑order modes is expected to yield even tighter constraints on extra‑dimensional parameters, a promising avenue that we leave for future exploration.

\section*{Acknowledgements}

We appreciate the helpful discussions with Dr. Xiao Zhi, Bo-Yang Zhang, and Yuan-Zhu Wang. This work is supported by the National Natural Science Foundation of China under Grants No.~12275238, No.~12542053, and No.~11675143, the National Key Research and Development Program of China under Grant No. 2020YFC2201503, the Zhejiang Provincial Natural Science Foundation of China under Grants No.~LR21A050001 and No.~LY20A050002, and the Fundamental Research Funds for the Provincial Universities of Zhejiang in China under Grant No.~RF-A2019015. Xin Zhang is supported by the National Natural Science Foundation of China (Grants No. 12473001 and No. 12533001), the National SKA Program of China (Grants No. 2022SKA0110200, and No. 2022SKA0110203), the China Manned Space Program (Grant No. CMS-CSST-2025-A02), and the 111 Project (Grant No. B16009).

\appendix

\end{document}